\documentstyle[12pt]{article}

\newcommand{\mt} {m_{top}}
\newcommand{\mh} {m_{Higgs}}
\newcommand{\mz} {M_{Z}}
\newcommand{\mw} {M_{W}}

\newcommand{\Z} {Z}
\newcommand{\ord} {\cal O}

\newcommand{\wz} {\Gamma_{Z}}

\newcommand{\wb} {\Gamma_{b}}

\def\hm{m_{_H}}
\def\Z0{Z^0}

\newcommand{\AmS}{{\protect\the\textfont2
  A\kern-.1667em\lower.5ex\hbox{M}\kern-.125emS}}

\begin{document}


\begin{center}

{\large THE TOP QUARK AND THE HIGGS BOSON MASS}
\vskip 4pt
{\large FROM LEP SLC AND CDF DATA}

\end{center}

\noindent
Guido MONTAGNA$^a$, Oreste NICROSINI$^b\footnote{\footnotesize
On leave from INFN, Sezione di Pavia}$, Giampiero PASSARINO$^c$
and Fulvio PICCININI$^d$ \\

\noindent
$^a$ INFN, Sezione di Pavia,  Italy \\
$^b$ CERN, TH Division, Geneva, Switzerland  \\
$^c$ Dipartimento di Fisica Teorica, Universit\`a di Torino and INFN,
Sezione di Torino, Italy \\
$^d$ Dipartimento di Fisica Nucleare e Teorica,
Universit\`a di Pavia, and INFN, Sezione di Pavia, Italy \\

\begin{abstract}
{\small The impact of the new experimental data from LEP, SLC and CDF on the
 top quark mass $m_{top}$ and the Higgs boson mass $m_{Higgs}$ is investigated.
The determinations of $m_{top}$ and of an upper bound on $m_{Higgs}$ are given,
taking into account the experimental error on the QED coupling constant
$\alpha_{em}$ and on the $b$ quark mass $m_b$. The relevance of higher order
theoretical uncertainties is pointed out. }
\end{abstract}

\begin{center}
Submitted to Physics Letters B
\end{center}



Up to now the four LEP experiments at CERN
collected roughly $8 \times 10^6$ $Z^0$
bosons, of which $3 \times 10^6$ have been produced in the 1993 resonance
scanning. This led to a substantial improvement in the measurement of the
$Z^0$ parameters such as $\mz$, $\wz$, $\wb$, the asymmetries and so
on~\cite{lepnew}.
Meanwhile other relevant experimental results have been achieved.
First, the experiment SLD at SLAC measured the value of the left-right
asymmetry on a sample of $5 \times 10^4$ $Z^0$'s, but with longitudinally
polarized electrons ($P \simeq 0.62$), reaching an accuracy competitive with
LEP determination~\cite{slac}.
Second, the experiment CDF at FERMILAB improved the
 measurement of $\mw$, leading to a better determination of the ratio
$\mw/\mz$~\cite{cdfmw}.
Last, but not least, very recently CDF collaboration claimed for
evidence of top quark production in $\bar p p $ collisions at $\sqrt{s} =
1.8$~TeV, quoting a value for the top quark mass of $\mt = 174 \pm
10^{+13}_{-12}$~GeV~\cite{cdfmt}.

At this point, it can be relevant to study the impact of the new experimental
data on the determination of the fundamental parameters of the Minimal
Standard Model, which is the goal of the present short note. Moreover, in the
light of the presently achieved experimental accuracy, two more items should
be taken into account. Firstly a particular care has
to be devoted to the effect of the experimental error on the electromagnetic
coupling constant $\alpha_{em}$, coming from the parameterization of the light
quark contribution to the vacuum polarization~\cite{vacpol},
and on the $b$ quark mass $m_b$.
Secondly the theoretical uncertainty due to higher order effects in the
Standard Model predictions has to be taken into account and properly
quantified.
It goes almost without saying that everything we are going to show
is largely based on data presented at the winter conferences;
in particular the averaging of $R_b$ among the four LEP experiments is
complicated and very preliminary so that this and other numbers, such as
the correlation matrix, could very well change~\cite{blond}.

In order to attain the above stated goal, the electroweak library of the code
 {\tt TOPAZ0}~\cite{TOPAZ0} has been used.
Very recent developments in the field of
electroweak and QCD radiative corrections, such as $\ord(G_F^2\mt^4)$ in
$\Delta\rho$, QCD corrections including $b$ quark mass effects with running
$b$ quark mass and full $\ord(\alpha_{em}\alpha_s)$ effects~\cite{lastew},
have been taken into account~\cite{TOPAZ02}.

The indirect determination of the top quark mass $\mt$ and the Higgs boson
mass $\mh$ have been studied in some detail. The data used are the
experimental measurements of the $Z^0$ parameters, namely $\mz$, $\Gamma_Z$,
$R$, $R_b$, $\sigma_{had}$, $g_V / g_A$ or the deconvoluted asymmetries, plus
the best determination of the ratio $\mw / \mz$ (UA2 + CDF, weighted average).
When the ratio $g_V / g_A $ has been used in place of the asymmetries,
the inclusion of the SLD measurement has been
performed by taking the weighted average of the LEP and SLD experimental data
(see Table~1).
The experimental error on $\alpha_{em} (\mz)$, $1/\alpha_{em}(light) =
128.87 \pm 0.12$, and on the $b$ quark mass $m_b$, $m_b = 4.7 \pm 0.2\,$GeV,
and the experimental value of the top quark mass $\mt$ as given by the direct
determination of CDF ($m_{top} = 174 \pm 17$~GeV)
have been included by proper penalty functions.
Moreover we have used the presently available elements of the correlation
matrix~\cite{priva}.

Let us begin with the top quark mass determination. The situation is well
summarized in Figs.~1-4, where $\chi^2$ versus $\mt$ is shown.
For a given $\mt$ the corresponding $\chi^2$ has been obtained by minimizing
the $\chi^2$ function with respect to $\mz$ and $\alpha_s$ for $\mh$ kept
fixed at $\mh = 300$~GeV ($\mz$ constrained at $91.190 \pm 0.004$~GeV, no
constraint on $\alpha_s$). Fig.~1 shows the $\chi^2$ in the following
situations ($\mw / \mz$ is always included):
LEP data only (dash-dotted line), LEP + SLC data (solid line),
LEP data + CDF
constraint (dotted line) and LEP + SLC data + CDF constraint (dashed line).
The ratio $g_V / g_A $ is used as summarizing the asymmetry data. The
uncertainty due to the error on $\alpha_{em}(\mz)$ and $m_b$ is propagated
in the theoretical part of the $\chi^2$.

Fig.~2 shows the same content as Fig.~1, but with $\alpha_{em}(\mz)$ and
$m_b$ kept fixed at their central value, 1/128.87 and 4.7~GeV respectively.
In Fig.~3 the effect of propagating the error on $\alpha_{em}(\mz)$ and
$m_b$ is pointed out by comparing a fit in which the error is taken into
account (dashed line) with a fit in which it is neglected (solid line).
Fig.~4, at
last, is the same as Fig.~1 but with the individual
asymmetries used in place of the combined value of $g_V / g_A$.
The best determination of $\mt$ can be considered the one in which the whole
set of experimental information is used, namely the one in which LEP + SLC
data  + CDF constraint (+ $\mw / \mz$) are used, together with the
propagation of the errors on $\alpha_{em}(\mz)$ and $m_b$.
In this case the best fit gives

\begin{equation}
\mt = 174.0^{+9.3+12.0+0.2}_{-9.6-12.5-3.4} \quad {\rm GeV} ,
\end{equation}
where, according to a commonly accepted procedure, the central value refers to
$\mh = 300\,$GeV,
the first error is statistical, the second one is obtained by allowing
$\mh$ to vary from 60 to 1000~GeV and the third one is due to higher order
theoretical uncertainties. At best fit one obtains $\alpha_s = 0.124$
and $\mz = 91.190$~GeV.
The last uncertainty is connected to the unknown electroweak
higher order terms, the truncation or not in perturbation theory,
the electroweak and QCD scales and the factorization or not of QCD radiation.
Actually the central value for $\mt$ deserves some additional explanation.
It has been derived by choosing some of the options on the treatment of
higher order EW terms such that we get the best agreement between {\tt
TOPAZ0}
and the other existing codes ({\tt BHM}~\cite{bhm},
{\tt LEPTOP}~\cite{leptop} and {\tt ZFITTER}~\cite{zfitter}).
If we use the same data set (LEP + SLC data + CDF constraint) and perform a
three parameter fit ($\mz, \mt, \alpha_s$) at $\mh$ fixed, then the
minimum of the $\chi^2$ corresponds to $\mh = 64$~GeV (more about this later,
see table 2). There is of course
some degree of arbitrariness in fixing $\mh$ to $300\,$GeV
and one could ask what happens if we derive results for $\mt$ at the best
value for $\mh$ kept fixed. Therefore we have performed a two parameter
fit with respect to $\mz$ and $\alpha_s$ for the Higgs mass fixed at
$64\,$GeV. We obtain

\begin{equation}
\mt = 161.9^{+9.4}_{-9.7} \quad {\rm GeV} ,
\end{equation}
corresponding to $\alpha_s = 0.122$. This result is confirmed by a three
parameter fit on the same data set, namely a fit to $\mz$, $\alpha_s$ and
$\mh$ (without any constraint on $\mh$), giving

\begin{equation}
\mt = 161.9^{+13.9}_{-11.4} \quad {\rm GeV} ,
\end{equation}
corresponding to $\alpha_s = 0.123$ and $\mh = 64$~GeV. Finally, by performing
the same type of fit with the penalty function on $\hm$, only
slightly different results are obtained, namely

\begin{equation}
\mt = 162.4^{+13.4}_{-9.6} \quad {\rm GeV} ,
\end{equation}
with $\alpha_s = 0.122$, $\mh = 68.5$~GeV. All these values are found to be in
good agreement with the results very recently obtained in~\cite{efl}.

Before making any comment it is worth noting that a slightly different
situation appears if we neglect the SLD data. Actually a
canonical fit at $\mh = 300$~GeV gives
\begin{equation}
\mt = 168.1^{+9.6+11.5}_{-9.9-11.8} \quad {\rm GeV} ,
\end{equation}
where the first error is statistical and the second one is due to a variation
of $\mh$ from 60 to 1000 GeV, whereas a fit in which $\mh $ is left free
provides
\begin{equation}
\mt = 164.0^{+14.7}_{-13.7} \quad {\rm GeV} ,
\end{equation}
with at best fit $\mh = 187$~GeV and $\alpha_s = 0.124$. The difference on
the central values for $\mt $ is smaller than the corresponding one appearing
when the SLD data is included, reflecting the fact that the SLD
asymmetry is about $3 \sigma $ away from the corresponding LEP measurement.
As a consequence of this the value of $\mh$ is driven towards the direct
search boundaries and the central value for $\mt$ depends strongly on the type
of fit performed. On the contrary we do not find large ($\approx 10\,$GeV)
deviations on $\mt$ from different fits if the SLD data is excluded.
At last excluding CDF constraint, i.e
 for the data set LEP + SLC (+$\mw / \mz$), the
best fit gives $\mt = 174.0^{+11.0+17.0+0.3}_{-11.7-18.5-4.9}$~GeV, in good
agreement with the result quoted in~\cite{bolek}. Moreover
for the LEP data alone (+$\mw / \mz$) we obtain
$\mt = 165^{+12+17}_{-13-19}$~GeV in agreement with~\cite{miquel}.

For the sake of comparison, it is worth
quoting the value of $\alpha_s$ as obtained from a fit to $R$, which gives

\begin{equation}
\alpha_s = 0.1258 \pm 0.0060^{+0.0029+0.0007}_{-0.0031-0.0014} ,
\end{equation}
where the first error is the experimental one, the second one comes from
$\mt = 174 \pm 17$~GeV and $\mh = 60 - 1000$~GeV and the last one is again
due to theoretical uncertainty. This value has been obtained along the same
lines of the one presented in~\cite{alphas}. If on the other hand we perform
a fit to $\mz, \mt, \mh$ to the LEP + SLC data + CDF constraint for $\alpha_s$
fixed and derive the $\chi^2(\alpha_s)$ distribution, then we get
$\alpha_s = 0.1218 \pm 0.0047$. The same fit excluding SLC gives instead
$\alpha_s = 0.1242^{+0.0053}_{-0.0050}$.

At this point some comments are in order. The SLC measurement of $A_{LR}$
increases
the fitted $\mt$ value of about 6-9~GeV with respect to the value given by
LEP data only. Moreover when the asymmetries are individually entered in the
fit instead of fitting the combined value of $g_V / g_A$,
the inclusion of the SLC measurement leads to a clear rise of the $\chi^2$.
This confirms that the SLC value is about 3~$\sigma$ away from the combined
LEP value of $g_V / g_A$. Including the CDF constraint
increases the fitted value of $\mt$ of about 3~GeV if SLC is not included,
whereas it gives no effect on the central value of $\mt$ if SLC is included
in the fit. In any case CDF constraint reduces the statistical error on $\mt$
of about 2~GeV and the error on $\mt$ due to the uncertainty on $\mh$ of
about 5~GeV. The uncertainty on the central value of $\mt$ generated by the
error on $\alpha_{em}(\mz)$ and $m_b$ can be quantified in about 2~GeV and
finally the one due to the theoretical ambiguity on higher orders can be
estimated to be around 4-5~GeV. It is also worth noting that the only $Z^0$
parameter which at present is {\it non-standard} is $R_b$, whose experimental
value is larger than expected of about two standard deviations, if indeed the
top quark is around $174\,$GeV. Excluding $R_b$ from the fit leads to an
increasing of $\mt$ of 4-6~GeV.

As far as $\mh$ determination is concerned, the $\chi^2$ as a function of
$\mh$ has been obtained by means of a three parameter fit with respect to
$\mz$, $\mt$ and $\alpha_s$ at $\mh$ fixed. In principle one could expect
some influence of the direct observation of the top quark on the theoretical
predictions for $\mh$. In order to point out such an effect the direct
determination of $\mt $ by CDF at $\mt = 174 \pm 17$~GeV has been taken into
account by including a proper penalty function. The situation is well described
by the results shown in Table~2 (7~observables means fitting $g_V / g_A$,
11~observables  means fitting the asymmetries). For the most complete
set of data (LEP + SLC + CDF), the curves at
95$\%$ C.L. in the $\mt$--$\mh$ plane are also shown in Fig.~5 for
three different values of $\alpha_s$ and including the Higgs
mass penalty function.

Predictions and corresponding
errors from a fit to LEP+SLD+CDF data (average $g_V/g_A$) are given in
Table~3, where $\sin^2 \vartheta (b)$ includes the universal $Z \to b \bar b$
vertex corrections.
The effect of the SLD measurement is to bring the $\mh$ upper limit well
below 1~TeV almost independently of the CDF constraint. The reason is that
SLD wants $\mt$ large and $\mh$ and $\alpha_s$ small in order to readjust
as much as possible the LR asymmetry.
The constraint on $\mh$ is more a symptom of the clash between SLD
and LEP than a reliable hint of $\mh $ small. The information
carried by the CDF constraint requires a careful examination. Actually it has
been verified that without the CDF constraint the $\chi^2 $ shape as a
function of
$\mh$ is unstable with respect to {\it normal} fluctuations of the experimental
data in the large $\mh$ tail, in agreement with~\cite{martinez}, whereas the
inclusion of the CDF constraint renders the tail more stable under small
perturbations of the data. In the case of $\mh$ determination the theoretical
uncertainty on EW higher orders plays a very relevant role. The situation is
described in Fig.~6, where the $\chi^2$ as a function of $\mh$ is plotted for
the most complete set of data LEP + SLD + CDF (7~observables).
Actually the $\chi^2 $ is not a
single curve but instead the whole band inside the two solid lines,
describing the theoretical uncertainty on the Standard Model
pseudo-observables.
Inside this band we have reported the $\chi^2$ distribution as derived from
{\tt TOPAZ0} in its default settings and also the one obtained
from {\tt TOPAZ0}
adapted for comparisons with other existing codes.
This theoretical uncertainty leads to a corresponding
uncertainty on the upper limit of some 200~GeV.

In conclusion, the last LEP, SLD and CDF data bring to an indirect
determination of $\mt$
at $\mt = 174^{+9.3+12.0+0.2}_{-9.6-12.5-3.4}$~GeV and allow to discuss an
upper limit on $\mh$ with some improvement with respect to the past.
\vskip 10pt

\noindent
Acknowledgments - The authors are grateful to Guido Altarelli for having
encouraged the present study, for several discussions on the subject and for a
critical reading of the preliminary manuscript.

\begin{table*}[htbp]\centering
\begin{tabular}{|c|c|}
\hline
Observable &    \\
\cline{2-2}
\hline
                     &                              \\
$\mz$                &    91.1899 $\pm$ 0.0044 GeV    \\
                     &                              \\
$\wz$                &    2497.1 $\pm$ 3.8     MeV    \\
                     &                              \\
$R$                  &    20.789 $\pm$ 0.04           \\
                     &                              \\
$R_b$                &    0.2210 $\pm$ 0.0019         \\
                     &                              \\
$\sigma_{had}$       &    51.51 $\pm$ 0.12     nb     \\
                     &                              \\
$A_{FB}^l$           &    0.0170 $\pm$ 0.0016         \\
                     &                              \\
$A_{pol}^{\tau}$     &    0.150  $\pm$ 0.010         \\
                     &                              \\
$A^e    $            &    0.120  $\pm$ 0.012         \\
                     &                              \\
$A_{FB}^b$           &    0.0970 $\pm$ 0.0045         \\
                     &                              \\
$A_{FB}^c$           &    0.072  $\pm$ 0.011         \\
                     &                              \\
$A_{LR} $            &    0.1668 $\pm$ 0.0079         \\
                     &                              \\
$g_V/g_A$(LEP)       &    0.0711 $\pm$ 0.0020         \\
                     &                              \\
$g_V/g_A$(LEP+SLD)   &    0.0737 $\pm$ 0.0018         \\
                     &                              \\
$\mw/\mz$            &    0.8814 $\pm$ 0.0021         \\
                     &                              \\
\hline

\end{tabular}
\caption[
Experimental values
]{\it
Experimental values
}
\label{ta1}
\end{table*}
\newpage

\begin{table*}[htbp]\centering
\begin{tabular}{|c|c|c|c|}
\hline
Set of data & $\chi^2_{min}$ & \multicolumn{2}{c|}
{$\mh$ (GeV)} \\
\cline{2-4}
            &         &   best value  &  95\% C.L.     \\
\hline
                           &           &      &       \\
LEP + SLD + CDF            &    7.5/7  &  64  &  580  \\
                           &           &      &       \\
LEP + CDF                  &    7.8/7  & 187  & 1354  \\
                           &           &      &       \\
LEP                        &    6.9/7  &  76  &  986  \\
                           &           &      &       \\
LEP + SLD                  &    6.6/7  &  39  &  400  \\
                           &           &      &       \\
LEP + SLD + CDF            &   17.1/11 &  53  &  511  \\
                           &           &      &       \\
LEP + CDF                  &    9.7/11 & 165  & 1237  \\
                           &           &      &       \\
\hline
\end{tabular}
\caption[
Predictions for $\mh$
]{\it
Predictions for $\mh$
}
\label{ta2}
\end{table*}
\newpage

\begin{table*}[htbp]\centering
\begin{tabular}{|c|c|c|c|c|}
\hline
Observable &  & stat. & Higgs & theor.  \\
\cline{2-5}
\hline
           &  &       &       &         \\
$M_W$         & $80.321$ GeV & $\pm 0.052$ & $\pm 0.019$ &
${}^{+0.004}_{-0.001}$  \\
           &  &       &       &         \\
$\sin^2\theta(l)$ & $0.2319$ & $\pm 0.0003$ & ${}^{+0.0002}_{-0.0004}$ &
${}^{+ \simeq 0}_{-0.0002}$ \\
           &  &       &       &         \\
$\sin^2\theta(b)$ & $0.2331$ & $\pm 0.0002$ &  $\pm 0.0005$ &
${}^{+ \simeq 0}_{-0.0002}$ \\
           &  &       &       &         \\
\hline
\end{tabular}
\caption[
Our predictions for $\mw$ and $\sin^2\theta(l,b)$ for a fit to
LEP+SLD+CDF data.
.]{\it
Our predictions for $\mw$ and $\sin^2\theta(l,b)$ for a fit to
LEP+SLD+CDF data.
}
\label{ta3}
\end{table*}
\newpage

\end{document}